\documentclass[twocolumn,showpacs,amsmath,amssymb,pra]{revtex4}
\usepackage{graphicx}
\usepackage{dcolumn}
\usepackage{bm}

\begin{document}

\title{Mutual neutralization in low energy $\text{H}^{+}+\text{H}^{-}$ collisions:\\
a quantum \textit{ab initio} study}

\author{Michael Stenrup}
\email[]{stenrup@physto.se}
\affiliation{Department of Theoretical Chemistry, School of Biotechnology, Royal Institute
of Technology, SE-106 91 Stockholm, Sweden}
\author{{\AA}sa Larson}
\author{Nils Elander}
\affiliation{Department of Physics, Stockholm University, AlbaNova University Center,
SE-106 91 Stockholm, Sweden}

\date{\today}

\begin{abstract}
The mutual neutralization of $\text{H}^+$ and $\text{H}^-$ at low collision energies
is studied by means of a molecular close-coupling approach. All degrees of freedom are
treated at the full quantum level also taking into account the identity of the nuclei.
The relevant $^1\Sigma_g^+$ and $^1\Sigma_u^+$ electronic states as well as the associated
non-adiabatic radial couplings are calculated for internuclear distances between 0.5 and
50~$a_0$. Following a transformation into a strictly diabatic basis, these quantities enter into
a  set of coupled equations for the motion of the nuclei. Numerical solution of these equations
allows the cross sections for neutralization into the $\text{H}(1)+\text{H}(n)$, $n=1,2,3$ final
states to be calculated. In the present paper, results are reported for the collision energy region
0.001 to 100~eV, with special emphasis on the important energy region below 10~eV. The low
temperature rate coefficient is obtained from a parametrization of the calculated cross
section and is estimated to be valid over the range 10 to 10,000~K.
\end{abstract}


\maketitle

\section{\label{level1:intro}INTRODUCTION}
The mutual neutralization of $\text{H}^+$ and $\text{H}^-$ is the prototype reaction for
electron transfer between two oppositely charged ions. With only two protons and two
electrons involved, it is the simplest ion-ion reaction possible. Here, calculations
should be able to provide accurate results which could serve as benchmarks in the
development of new theoretical methods and models. Furthermore, the reactants have well
defined quantum states and are relatively easy to prepare in the laboratory.
A detailed comparison between experiment and theory should thus be possible and should
provide a natural starting point to understanding more complicated ion-ion collision
phenomena. However, the simplicity of this reaction is deceptive, a fact which has
stimulated a large number of studies in the past. Nevertheless, surprisingly few of these
have investigated the neutralization cross section for collision energies below a few eV,
which plays an important role in determining the $\text{H}^-$ abundance in low temperature
ionized environments. A compelling example from astrophysics is its influence on the gas
phase formation of $\text{H}_2$, in particular in the chemistry of the primordial gas,
where one of the main reaction paths involves $\text{H}^-$ as an
intermediate~\cite{gal98,glo03,glo06}.

Bates and Lewis~\cite{bat55} were the first to study the mutual neutralization of
$\text{H}^+$ and $\text{H}^-$, which is written schematically as
\begin{equation}
\label{eq:mnr}
\text{H}^++\text{H}^-\rightarrow\text{H}(1)+\text{H}(n)~,\nonumber
\end{equation}
where $n$ is the main quantum number of the excited hydrogen atom. Their calculations were
based on the semiclassical Landau-Zener model~\cite{lan32,zen32} in which the reaction is
thought to proceed along a system of horizontal covalent potential energy curves crossed by
an attractive ion-pair potential. This pioneering study was later followed by others,
among them by Olson \textit{et al.}~\cite{ols70} and more recently by Eerden
\textit{et al.}~\cite{eer95}. Fussen and Kubach~\cite{fus86} have gone beyond the
Landau-Zener concept and investigated the reaction at low collision energies using a
one electron close-coupling model.

The only measurement of the neutralization cross-section for collision energies below 3~eV
is the merged beam experiment of Moseley \textit{et al.}~\cite{mos70}. Their results seem
to indicate a cross section that is approximately a factor of three higher than obtained in
most of the cited theoretical studies. However, the theoretical results are
mutually consistent and, for higher energies, also in more or less good agreement with the
measurements reported by Szucs \textit{et al.}~\cite{szu84} and Peart and Hayton~\cite{pea92}.
This intriguing situation has led to the suspicion that the low energy cross section of
Moseley \textit{et al.} is overestimated~\cite{szu84,pea85,pea89,pea92}.

In this paper we report results of a fully quantum mechanical \textit{ab initio} study of the
reaction at hand. We are mainly concerned with the collision energy region 0.001 to 10~eV,
but the calculations have been extended up to 100~eV in order to facilitate comparison with
the cited experiments. The present study is based on a molecular close-coupling expansion of
the total wave function. Accurate electronic structure methods have been used to calculate
adiabatic states and non-adiabatic couplings of both gerade and ungerade symmetry over a wide
range of internuclear distances. Subsequently, a transformation into a strictly diabatic
representation has been performed and the diabatic quantities entered into a set of
coupled equations for the motion of the nuclei. By numerically solving these equations, we have
been able to calculate the neutralization cross sections for scattering within both the
gerade and ungerade inversion symmetries. Finally, the identity of the nuclei has been
accounted for by properly combining the gerade and ungerade cross sections.

The paper is organized as follows. Section~\ref{level1:theory} describes the theoretical
framework underlying the present study. Section~\ref{level1:esc} describes the electronic
structure calculations and discusses the resulting potential energy curves and coupling
matrix elements. Section~\ref{level1:numpro} gives a brief account of the numerical
treatment of the scattering problem. Section~\ref{level1:res} presents and discusses the
calculated neutralization cross sections and the reaction rate coefficient.
Section~\ref{level1:sum} concludes the paper and summarizes the main results. Unless stated
otherwise, vector and matrix quantities are written in boldface and atomic units are used.

\section{\label{level1:theory}THEORY}

\subsection{\label{level2:molexp}The molecular expansion: adiabatic and diabatic
formulations}
Consider a diatomic system composed of two nuclei and $N$ electrons. The masses and charges
of the nuclei are denoted by $M_i$ and $Z_i$, respectively. Let
$\mathbf{R}=(R,\theta,\varphi)$ be the internuclear separation vector and $\mathbf{r}_i$ a
vector pointing from the center of mass of the nuclei to the \textit{i}-th electron. All
vectors are expressed with respect to a space fixed axis system. For convenience, the
notation $\mathbf{r}=(\mathbf{r}_1,\mathbf{r}_2,\ldots,\mathbf{r}_N)$ will occasionally be
used. In these coordinates, the center of mass motion of the total system can be separated
out, and the Hamiltonian takes the form~\cite{gel97}
\begin{equation}
\label{eq:ham}
H=-\frac{1}{2\mu}\nabla_\mathbf{R}^2+H^{\text{el}}~,
\end{equation}
where $\mu$ is the reduced mass of the nuclei and $H^{\text{el}}$ is the electronic
Hamiltonian. Neglecting terms of the form
$[2(M_1+M_2)]^{-1}\nabla_{\mathbf{r}_i}\cdot\nabla_{\mathbf{r}_j}$, it follows that
\begin{eqnarray}
\label{eq:elham}
H^\text{el}&=&-\frac{1}{2}\nabla_\mathbf{r}^2+\frac{Z_1Z_2}{R}
+\sum_{i>j=1}^N\frac{1}{\left|\mathbf{r}_i-\mathbf{r}_j\right|}\nonumber\\
&&-\sum_{i=1}^N\left(\frac{Z_1}{\left|\frac{\mu}{M_1}\mathbf{R}+\mathbf{r}_i\right|}
+\frac{Z_2}{\left|\frac{\mu}{M_2}\mathbf{R}-\mathbf{r}_i\right|}\right)~.
\end{eqnarray}

The total wave function $\psi(\mathbf{R},\mathbf{r})$ satisfies the time-independent
Schr{\"o}dinger equation
\begin{equation}
\label{eq:se}
H\psi(\mathbf{R},\mathbf{r})=E\psi(\mathbf{R},\mathbf{r})~,
\end{equation}
where $E$ is the total energy in the center of mass system. We expand
$\psi(\mathbf{R},\mathbf{r})$ in terms of electronic and nuclear states,
$\phi_i^\text{a}(\mathbf{R},\mathbf{r})$ and $\chi_i^\text{a}(\mathbf{R})$, according to
\begin{equation}
\label{eq:molexp}
\psi(\mathbf{R},\mathbf{r})=
\sum_{i=1}^\infty\phi_i^\text{a}(\mathbf{R},\mathbf{r})\chi_i^\text{a}(\mathbf{R})
=\bm{\phi}^\text{a}\bm{\chi}^\text{a}~,
\end{equation}
where $\phi_i^\text{a}(\mathbf{R},\mathbf{r})$ are normalized solutions to the
electronic Schr{\"o}dinger equation
\begin{equation}
\label{eq:else}
H^\text{el}\phi_i^\text{a}(\mathbf{R},\mathbf{r})=
\epsilon_i^\text{a}(R)\phi_i^\text{a}(\mathbf{R},\mathbf{r})~.
\end{equation}
The electronic energies $\epsilon_i^\text{a}(R)$ depend parametrically on~$R$ and obey
the non-crossing rule~\cite{neu29}. The particular choice of basis functions defined by
Eq.~(\ref{eq:else}) is called the adiabatic representation and is denoted by the
superscript~a. Inserting the expansion~(\ref{eq:molexp}) into Eq.~(\ref{eq:se}) and using
the orthonormality of the electronic states yields a set of coupled differential equations
for the motion of the nuclei:
\begin{eqnarray}
\label{eq:adse}
\left(-\frac{1}{2\mu}\mathbf{1}\nabla_\mathbf{R}^2
-\frac{1}{\mu}\mathbf{F}^\text{a}\cdot\nabla_\mathbf{R}
-\frac{1}{2\mu}\mathbf{G}^\text{a}
+\mathbf{V}^\text{a}\right)\bm{\chi}^\text{a}\nonumber\\
=E\mathbf{1}\bm{\chi}^\text{a}~, 
\end{eqnarray}
with
\begin{equation}
\label{eq:fdnac}
\mathbf{F}_{ij}^\text{a}(\mathbf{R})=
\left\langle\phi_i^\text{a}\left|\nabla_\mathbf{R}\right|\phi_j^\text{a}\right\rangle~,
\end{equation}
\begin{equation}
\label{eq:sdnac}
G_{ij}^\text{a}(\mathbf{R})=
\left\langle\phi_i^\text{a}\left|\nabla_\mathbf{R}^2\right|\phi_j^\text{a}\right\rangle~,
\end{equation}
\begin{equation}
\label{eq:adpot}
V_{ij}^\text{a}(R)=
\left\langle\phi_i^\text{a}\left|H^{\text{el}}\right|\phi_j^\text{a}\right\rangle~.
\end{equation}
By definition, the matrix $\mathbf{V}^\text{a}$ is diagonal and contains the adiabatic
potential energy curves $\epsilon_i^\text{a}(R)$. The matrix vector $\mathbf{F}^\text{a}$
and the matrix $\mathbf{G}^\text{a}$ both have off-diagonal elements, which are the
non-adiabatic couplings between different electronic states. In addition,
$\mathbf{G}^\text{a}$ also has diagonal elements commonly referred to as adiabatic energy
corrections. If the electronic wave functions are taken to be real, the chain rule for
derivatives can be used to show that $\mathbf{F}^\text{a}$ is antisymmetric, i.e.,
\begin{equation}
\label{eq:asc}
\mathbf{F}_{ij}^\text{a}(\mathbf{R})=-\mathbf{F}_{ji}^\text{a}(\mathbf{R})~.
\end{equation}			

The electronic states approximately correlate with the separated atomic limits and therefore
each term in Eq.~(\ref{eq:molexp}) defines a scattering channel. To determine the outcome of
any collision process, we need to solve the coupled equations out to the asymptotic region
where these limits are reached. However, it has been known for a long time that a product
of the form $\chi_i^\text{a}(\mathbf{R})\phi_i^\text{a}(\mathbf{R},\mathbf{r})$ does not
properly describe the translational motion of the electrons along with the
nuclei~\cite{bat58,del81,bel01}. This can lead to difficulties of both formal and practical
nature. One of the most prominent is the appearance of non-vanishing couplings at large or
even infinite internuclear distances~\cite{jep60,del81,bel01}. However, it is demonstrated
in Sec.~\ref{level2:tcs} that the influence of these couplings on the present results is
expected to be small.

The non-adiabatic coupling terms in Eq.~(\ref{eq:adse}) often vary rapidly with $R$ and may
cause problems in the numerical solution procedure. Furthermore, the presence of
$\mathbf{F}^\text{a}$ leads to differential equations which contain both first and second
order derivatives. It is therefore desirable to transform Eq.~(\ref{eq:adse}) into an
electronic basis where some or all of the non-adiabatic couplings disappear. Following Mead
and Truhlar~\cite{mea82}, we shall refer to a representation in which all three components
of $\mathbf{F}^\text{a}$ vanish as strictly diabatic. From this point on, it is assumed that
the total wave function can be represented in a finite set of $M$ adiabatic basis
functions. The transformation may then be written in terms of an orthogonal
matrix~$T_{ij}(\mathbf{R})$~\cite{bae02} as
\begin{equation}
\label{eq:atdt1}
\phi_i^\text{d}(\mathbf{R},\mathbf{r})=
\sum_{j=1}^M\phi_j^\text{a}(\mathbf{R},\mathbf{r})T_{ij}^\text{T}(\mathbf{R})~.
\end{equation} 
In order for the total wave function to be preserved, the nuclear states must transform
according to
\begin{equation}
\label{eq:atdt2}
\chi_i^\text{d}(\mathbf{R})=
\sum_{j=1}^M\chi_j^\text{a}(\mathbf{R})T_{ij}^\text{T}(\mathbf{R})~.
\end{equation}
Using the relation~(\ref{eq:atdt2}) to replace the adiabatic states in Eq.~(\ref{eq:adse})
yields after some manipulation
\begin{eqnarray}
\label{dise1}
\left(-\frac{1}{2\mu}\mathbf{1}\nabla_\mathbf{R}^2
-\frac{1}{\mu}\mathbf{F}^\text{d}\cdot\nabla_\mathbf{R}
-\frac{1}{2\mu}\mathbf{G}^\text{d}+\mathbf{V}^\text{d}\right)\bm{\chi}^\text{d}\nonumber\\
=E\mathbf{1}\bm{\chi}^\text{d}~,
\end{eqnarray}
with
\begin{equation}
\mathbf{F}^\text{d}=\mathbf{T}^\text{T}\left(\nabla_\mathbf{R}
+\mathbf{F}^\text{a}\right)\mathbf{T}~,
\end{equation}
\begin{equation}
\mathbf{G}^\text{d}=\mathbf{T}^\text{T}\left(\nabla_\mathbf{R}^2
+2\mathbf{F}^\text{a}\cdot\nabla_\mathbf{R}
+\mathbf{G}^\text{a}\right)\mathbf{T}~,
\end{equation}
\begin{equation}
\mathbf{V}^\text{d}=\mathbf{T}^\text{T}\mathbf{V}^\text{a}\mathbf{T}~.
\end{equation}
The first derivative coupling $\mathbf{F}^\text{d}$ expressed in the new basis is seen to
vanish provided that $\mathbf{T}$ is a solution to the equation 
\begin{equation}
\label{eq:atdte1}
\left(\nabla_\mathbf{R}+\mathbf{F}^\text{a}\right)\mathbf{T}=\mathbf{0}~. 
\end{equation}
In the special case where all electronic states have the same angular momenta projection
onto the molecular axis, only the radial component of $\mathbf{F}^\text{a}$ can be
non-zero~\cite{mea82}. Equation~(\ref{eq:atdte1}) then reduces to
\begin{equation}
\label{eq:atdte2}
\left[\frac{\text{d}}{\text{d}R}+\bm{\tau}(R)\right]\mathbf{T}(R)=\mathbf{0}~,
\end{equation}
where
\begin{equation}
\label{eq:radnac}
\tau_{ij}(R)=\left\langle\phi_i^\text{a}\left|\frac{\partial}{{\partial}R}
\right|\phi_j^\text{a}\right\rangle~.
\end{equation}
Assuming that the first derivative couplings between the adiabatic states with
$i{\leq}M$ and those with $i>M$ are zero, also $\mathbf{G}^\text{d}$ can be shown to
vanish~\cite{bae02}. The coupled equations then take the form
\begin{eqnarray}
\label{eq:dise2}
\left(-\frac{1}{2\mu}\mathbf{1}\nabla_\mathbf{R}^2+\mathbf{V}^\text{d}\right)
\bm{\chi}^\text{d}=E\mathbf{1}\bm{\chi}^\text{d}~.
\end{eqnarray}

In the strictly diabatic representation, the non-adiabatic couplings transform into
off-diagonal elements of the potential matrix $\mathbf{V}^\text{d}$ (electronic couplings)
and as a consequence the diagonal elements of this matrix (diabatic potentials) no longer
obey the non-crossing rule. As will be illustrated in Sec.~\ref{level2:atdt}, the strictly
diabatic representation may show little resemblance to the simple and intuitive curve
crossing picture often referred to as diabatic.    

\subsection{\label{level2:abcics}Asymptotic boundary conditions and the integral
cross section}
The interpretation of $\chi_i^\text{d}(\mathbf{R})$ is usually simplified by requiring that
\begin{eqnarray}
\label{eq:atdtbc}
\lim_{R\rightarrow\infty}\mathbf{T}(R)=\mathbf{1}~,
\end{eqnarray}
which makes the adiabatic and the diabatic representations coincide in the region where
the scattering wave function is evaluated. Thus, in any of the representations, the
physical situation is that of an incoming plane wave in some entrance channel~$j$, and
outgoing spherical waves in all channels which are energetically allowed. The direction of
the incoming plane wave can be chosen along the space fixed z-axis. If the potential falls
off reasonably fast~\cite{mot65}, the appropriate boundary conditions are given by
\begin{equation}
\label{eq:abc}
\chi_i^\text{d}(\mathbf{R})
\underset{R\rightarrow\infty}{\sim}\delta_{ij}e^{ik_jz}+f_{ij}(E,\theta)\frac{e^{ik_iR}}{R}~,
\end{equation}
where $f_{ij}(E,\theta)$ is the scattering amplitude for entering the collision in channel~$j$
and leaving it in channel~$i$. Since the potential matrix is spherically symmetric, the
scattering amplitude is independent of the azimuthal angle~$\varphi$. The asymptotic wave
number~$k_i$ is defined as
\begin{equation}
\label{eq:k1}
k_i=\left[2\mu\left(E-E_i^\text{th}\right)\right]^{1/2}~,
\end{equation}
where $E_i^\text{th}=\lim_{R\rightarrow\infty}V_{ii}^\text{d}(R)$ is the corresponding
threshold energy. In the case of a Coulomb potential, the exponents in Eq.~(\ref{eq:abc})
have to be modified with logarithmic phase factors which arise due to the long range nature
of this interaction~\cite{mot65}.

Evaluating the probability fluxes associated with the two terms in Eq.~(\ref{eq:abc}) leads
to the familiar multichannel expression for the integral cross section:
\begin{equation}
\label{eq:totcs1}
\sigma_{ij}(E)=
\frac{2\pi{k_i}}{k_j}\int_{0}^{\pi}\left|f_{ij}(E,\theta)\right|^2\sin{\theta}\text{d}\theta~.
\end{equation}

If the adiabatic states are connected by non-vanishing couplings at infinity, neither the
boundary condition~(\ref{eq:atdtbc}) nor the scattering formalism developed below will be
appropriate. In this case, the couplings can be manually cut at some large but finite
internuclear distance. This procedure is usually justified if the collision energy is low
enough, but its validity should be considered from case to case.

\subsection{\label{level2:scafor}The scattering formalism}
The conventional approach to solving Eq.~(\ref{eq:dise2}) is to express
$\bm{\chi}^\text{d}(\mathbf{R})$ in terms of a partial wave expansion,
\begin{equation}
\label{eq:pwe}
\chi_i^\text{d}(\mathbf{R})=\frac{1}{R}\sum_{l=0}^{\infty}A_lu_{i,l}(R)P_l(\cos\theta)~,
\end{equation}
where $P_l(\cos\theta)$ are the well known Legendre polynomials \cite{abr72} and $A_l$ are
constants to be chosen so that the boundary conditions (\ref{eq:abc}) are fulfilled. The
radial wave functions $u_{i,l}(R)$ can be shown to satisfy the equations
\begin{eqnarray}
\label{eq:rse1}
\left[-\frac{1}{2\mu}\frac{\text{d}^2}{\text{d}R^2}+\frac{l(l+1)}{2\mu{R^2}}\right]u_{i,l}(R)
+\sum_{j=1}^MV_{ij}^\text{d}u_{j,l}(R)\nonumber\\
=Eu_{i,l}(R)~,
\end{eqnarray}
subject to the physical boundary conditions $u_{i,l}(0)=0$. If the corresponding derivatives
$u_{i,l}^\prime(0)$ are chosen appropriately, these equations have $M$ linearly independent
solutions which can be combined into a square matrix~$\widetilde{\mathbf{u}}$. It follows
that
\begin{equation}
\label{eq:rse2} 
\left(\frac{\text{d}^2}{\text{d}R^2}\mathbf{1}+\mathbf{Q}_l\right)\widetilde{\mathbf{u}}_l
=\mathbf{0}~,
\end{equation}
where
\begin{equation}
\label{eq:q}
\mathbf{Q}_l=2\mu\left(E\mathbf{1}-\mathbf{V}^\text{d}\right)-\frac{l(l+1)}{R^2}\mathbf{1}~.
\end{equation}

It is now assumed that the potential matrix is diagonal beyond some radius $\rho$ and here has
elements corresponding to either pure Coulomb or pure short-range interactions. In
this region, Eq.~(\ref{eq:rse2}) is analytically solvable and the radial wave functions
may be expressed in terms of the incoming and outgoing wave solutions, $\alpha_{ij,l}(R)$
and $\beta_{ij,l}(R)$, as defined in the Appendix. These are essentially the same as
adopted by Johnson~\cite{joh85} but have been extended to cover the Coulomb case. Thus,
\begin{equation}
\label{eq:sdef}
\widetilde{\mathbf{u}}_l(R)\underset{R\ge\rho}{=}
\bm{\alpha}_l(R)-\bm{\beta}_l(R)\mathbf{S}_l~,
\end{equation}
which defines the scattering matrix $\mathbf{S}_l$. Inserting this expression into the
partial wave expansion~(\ref{eq:pwe}) and comparing with the boundary
conditions~(\ref{eq:abc}) leads to the well known formula
\begin{equation}
\label{eq:scaamp}
f_{ij}(E,\theta)=\frac{1}{2i(k_ik_j)^{1/2}}
\sum_{l=0}^{\infty}(2l+1)\left(S_{ij,l}^\text{oo}-\delta_{ij}\right)P_l(\cos\theta),
\end{equation}
for the scattering amplitude in terms of the open-open or physical partition of
$\mathbf{S}_l$. The integral cross section for scattering from channel~$j$ to channel~$i$
is readily evaluated to be
\begin{equation}
\label{eq:totcs2}
\sigma_{ij}(E)=\sum_{l=0}^{\infty}\sigma_{ij,l}(E)~,
\end{equation}
where
\begin{equation}
\label{eq:partcs}
\sigma_{ij,l}(E)=\frac{\pi}{k_j^2}(2l+1)\left|S_{ij,l}^\text{oo}-\delta_{ij}\right|^2~.
\end{equation}

Instead of working with $\mathbf{S}_l$ directly, it is convenient to introduce the reactance
matrix~$\mathbf{K}_l$~\cite{gol64} which is defined by
\begin{equation}
\label{eq:kdef}
\widetilde{\mathbf{u}}_l(R)\underset{R\ge\rho}{=}
\mathbf{a}_l(R)-\mathbf{b}_l(R)\mathbf{K}_l~,
\end{equation}
where $a_{ij,l}(R)$ and $b_{ij,l}(R)$ are regular and irregular solutions to
Eq.~(\ref{eq:rse2}) explicitly given in the Appendix. Combining the definitions above
together with those from the Appendix leads to the relationship~\cite{joh85}
\begin{equation}
\label{eq:cayley}
\mathbf{S}_l^\text{oo} = (1+i\mathbf{K}_l^\text{oo})^{-1}(1-i\mathbf{K}_l^\text{oo})~,  
\end{equation}
between the open-open partitions of $\mathbf{S}_l$ and $\mathbf{K}_l$. This expression
resembles the Cayley transform and, since $\mathbf{S}_l^\text{oo}$ is both symmetric and
unitary, implies that $\mathbf{K}_l^\text{oo}$ is real~\cite{mot65}.

Finally, we introduce the logarithmic derivative
\begin{equation}
\label{eq:ydef}
\mathbf{y}_l=\widetilde{\mathbf{u}}_l^\prime\widetilde{\mathbf{u}}_l^{-1}~,
\end{equation}
which transforms the radial Schr{\"o}dinger equation into the numerically more stable
matrix Riccati equation~\cite{joh73,man93}
\begin{equation}
\label{eq:riceq}
\mathbf{y}_l^\prime+\mathbf{Q}_l+\mathbf{y}_l^2=\mathbf{0}~,
\end{equation}
while the boundary condition at the origin becomes that of a diagonal matrix with infinite
elements. This is the set of coupled equations that we ultimately have to solve. The
relationship between $\mathbf{y}_l$ and $\mathbf{K}_l$ is easily seen to be
\begin{eqnarray}
\label{eq:kyrel}
\mathbf{K}_l&=&
\left[\mathbf{y}_l(\rho)\mathbf{b}_l(\rho)-\mathbf{b}_l^\prime(\rho)\right]^{-1}\nonumber\\
&&\times\left[\mathbf{y}_l(\rho)\mathbf{a}_l(\rho)-\mathbf{a}_l^\prime(\rho)\right]~.
\end{eqnarray}
         
\subsection{\label{level2:idnuc}Modifications in the case of identical nuclei}
Up to this point in the discussion complications that might arise if the nuclei are
identical have been ignored. These may be considered in two steps. Firstly, since the
charges of the nuclei that enter $H^\text{el}$ are equal, the electronic states will be either
symmetric (gerade) or antisymmetric (ungerade) under inversion of $\mathbf{r}$. If the
nuclei are labeled A and B, and $\alpha,\beta,\gamma,...$ represent groups of electrons in
some definite quantum states, each asymptotic channel will be a linear combination of two
configurations, say $\text{A}(\alpha)+\text{B}(\beta)$ and $\text{A}(\beta)+\text{B}(\alpha)$.
Such linear combinations can not describe the situation where a nucleus is known to carry
a certain number of electrons in a certain quantum state (unless $\alpha=\beta$).
Localization of the electron cloud can be achieved by forming proper linear combinations
of the gerade and ungerade solutions. The $g$/$u$ symmetry is explicitly
indicated as a superscript and the channel index $i$ is allowed to include all other
quantum numbers. The appropriate scattering amplitudes are then given by~\cite{mot65}
\begin{equation}
\label{eq:diamp}
f_{ij}^\text{di}(E,\theta)=
\frac{1}{2}\left[f_{ij}^g(E,\theta)+f_{ij}^u(E,\theta)\right]
\end{equation}
and
\begin{equation}
\label{eq:examp}
f_{ij}^\text{ex}(E,\theta)=
\frac{1}{2}\left[f_{ij}^g(E,\theta)-f_{ij}^u(E,\theta)\right]~,
\end{equation}
corresponding to the direct and exchange reactions
\begin{equation}
\label{eq:dipro}
\text{A}(\alpha)+\text{B}(\beta)\rightarrow\text{A}(\gamma)+\text{B}(\delta)
\end{equation}
and
\begin{equation}
\label{ex:expro}
\text{A}(\alpha)+\text{B}(\beta)\rightarrow\text{A}(\delta)+\text{B}(\gamma)~.
\end{equation}
Here, the amplitudes $f_{ij}^g$ and $f_{ij}^u$ are given by the usual
formula~(\ref{eq:scaamp}) applied to the gerade and ungerade manifolds separately.
      
A second complication arises from the fact that also the masses of the nuclei are equal.
Consequently, it is impossible to distinguish direct scattering in the direction $\theta$
from exchange scattering in the direction $\pi-\theta$. This effect can be accounted for by
adding the direct and exchange amplitudes coherently. The fully symmetrized scattering
amplitudes are~\cite{mas69}
\begin{equation}
\label{eq:symamp}
f_{ij}^{\pm}(E,\theta)=f_{ij}^\text{di}(E,\theta)\pm f_{ij}^\text{ex}(E,\pi-\theta)~,
\end{equation}
where the sign depends on whether the spatial part of the wave function should be
symmetric~(+) or antisymmetric~(-) under exchange of the nuclei. To illustrate this point
we consider a hydrogen quasi molecule where the two protons are known to be in a singlet
spin state. According to the Pauli principle, the total wave function (spin part included)
must be antisymmetric under exchange of the nuclei. Since the singlet spin state itself
is antisymmetric under such an operation, the spatial part of the wave function is forced
to be symmetric. Thus, $f_{ij}^+$ is the appropriate choice of amplitude.
            
The integral cross section is obtained by taking the modulus square of Eq.~(\ref{eq:symamp})
and integrating the result over the unit sphere: 
\begin{equation}
\label{eq:symcs}
\sigma_{ij}^+(E)
=\sum_{l~\text{even}}\sigma_{ij,l}^g(E)+\sum_{l~\text{odd}}\sigma_{ij,l}^u(E)
\end{equation}
and
\begin{equation}
\label{eq:asymcs}
\sigma_{ij}^-(E)
=\sum_{l~\text{odd}}\sigma_{ij,l}^g(E)+\sum_{l~\text{even}}\sigma_{ij,l}^u(E)~,
\end{equation}
where $\sigma_{ij,l}^g$ and $\sigma_{ij,l}^u$ are given by
Eq.~(\ref{eq:partcs}) applied to the gerade and ungerade manifolds separately. The
subscript even (odd) indicates that only even (odd) partial waves should be summed over.
In the particular case of the hydrogen quasi molecule, the total spin state of the nuclei
is either a singlet (para hydrogen) or a triplet (ortho hydrogen). Taking into account the
degeneracy factors (1 and 3) and the exchange symmetry of these spin states, the
symmetrized integral cross section for an ensemble of particles with randomly oriented
spins is
\begin{eqnarray}
\label{eq:h2cs}
\sigma_{ij}^\text{sym}(E)&=&\frac{1}{4}\sum_{l~\text{even}}\sigma_{ij,l}^g(E)
+\frac{3}{4}\sum_{l~\text{odd}}\sigma_{ij,l}^g(E)\nonumber\\
&&+\frac{3}{4}\sum_{l~\text{even}}\sigma_{ij,l}^u(E)
+\frac{1}{4}\sum_{l~\text{odd}}\sigma_{ij,l}^u(E)~.\nonumber\\
\end{eqnarray}
This can be compared with the expression
\begin{eqnarray}
\label{eq:h2cs2}
\sigma_{ij}^\text{dist}(E)&=&\frac{1}{2}\sum_{l=0}^\infty\sigma_{ij,l}^g(E)
+\frac{1}{2}\sum_{l=0}^\infty\sigma_{ij,l}^u(E)~,
\end{eqnarray}
obtained by treating the nuclei as distinguishable, and thus adding the direct and exchange
amplitudes incoherently.

In deriving Eqs.~(\ref{eq:h2cs}) and (\ref{eq:h2cs2}) we have assumed that all final
states exist in both gerade and ungerade versions. This is not always the case. Of particular
interest to the present study are processes in which the final electronic state corresponds to
$\gamma=\delta$ and exists only in a gerade version. In this case, Eqs.~(\ref{eq:h2cs}) and
(\ref{eq:h2cs2}) are still valid (setting $\sigma_{ij,l}^u=0$), although the arguments leading
to these formulas are somewhat modified.         

\section{\label{level1:esc}ELECTRONIC STRUCTURE CALCULATIONS}
The electronic states relevant to the present study are those of $\text{H}_2$. The amount
of data published for this system is of course substantial. Of particular relevance here
are the very accurate electronic structure calculations reported by Wolniewicz and
co-workers~\cite{wol88,wol90,wol93,wol94,wol95,wol98,wol02} using explicitly correlated
basis functions and the calculations by Detmer \textit{et al.}~\cite{det98} using a
non-spherical Gaussian basis set. Reactions like the present one, however, require the
calculation of not only a large number of excited states but also the associated
non-adiabatic couplings for a wide range of internuclear distances. To the best of our
knowledge, a complete and accurate set of data for these quantities does not exist in the
literature. We have therefore performed \textit{ab initio} electronic structure
calculations of all the relevant potential energy curves and non-adiabatic couplings.
Where such data is available, our results have been compared with the benchmark
calculations of Refs.~\cite{wol88,wol90,wol93,wol94,wol95,wol98,wol02}.
 
\subsection{\label{level2:pes}Potential energy curves}
We have calculated the adiabatic potential energy curves and the associated electronic wave
functions corresponding to the seven lowest $^1\Sigma_g^+$ states and the six
lowest $^1\Sigma_u^+$ states, here denoted as (1-7)$^1\Sigma_g^+$ and
(1-6)$^1\Sigma_u^+$, respectively. All calculations have been performed at the
full configuration interaction (FCI) level using a modified version of the Dalton~2.0
program package~\cite{dal05}. The molecular orbitals have been obtained in a
(11\textit{s},8\textit{p},7\textit{d},2\textit{f}) spherical Gaussian basis set contracted
to [9\textit{s},8\textit{p},7\textit{d},2\textit{f}].  The compact basis functions of this
set have been taken from the augmented correlation-consistent polarized valence quadruple
zeta (aug-cc-pVQZ) basis set of Dunning and co-workers~\cite{dun89,ken92}, while the more
diffuse ones have been approximately optimized to obtain a good representation of the
excited states. The basis set is given in Table~\ref{table:basset}.
\begin{table}
\caption{\label{table:basset}Specification of the spherical Gaussian basis set used in the
present study.}
\begin{ruledtabular}
\begin{tabular}{lccr}
Type & Exponent & Type & Exponent \\
\hline
\textit{s}\footnote{Contraction of six primitive \textit{s} functions with exponents
(82.64, 12.41, 2.824, 0.7977, 0.2581, 0.08989) and coefficients
(0.002006, 0.015343, 0.075579, 0.256875, 0.497368, 0.296133).} 
&-&\textit{p}&0.0275700\\
\textit{s}&0.7977000&\textit{p}&0.0094220\\
\textit{s}&0.2581000&\textit{p}&0.0031910\\
\textit{s}&0.0898900&\textit{p}&0.0010470\\
\textit{s}&0.0251300&\textit{d}&2.0620000\\
\textit{s}&0.0078770&\textit{d}&0.6620000\\
\textit{s}&0.0029256&\textit{d}&0.1900000\\
\textit{s}&0.0008852&\textit{d}&0.0635300\\
\textit{s}&0.0002917&\textit{d}&0.0211100\\
\textit{p}&2.2920000&\textit{d}&0.0070570\\
\textit{p}&0.8380000&\textit{d}&0.0023460\\
\textit{p}&0.2920000&\textit{f}&1.3970000\\
\textit{p}&0.0848000&\textit{f}&0.3600000\\
\end{tabular}
\end{ruledtabular}
\end{table}
\begin{figure}[htp]
\scalebox{0.68}{\includegraphics{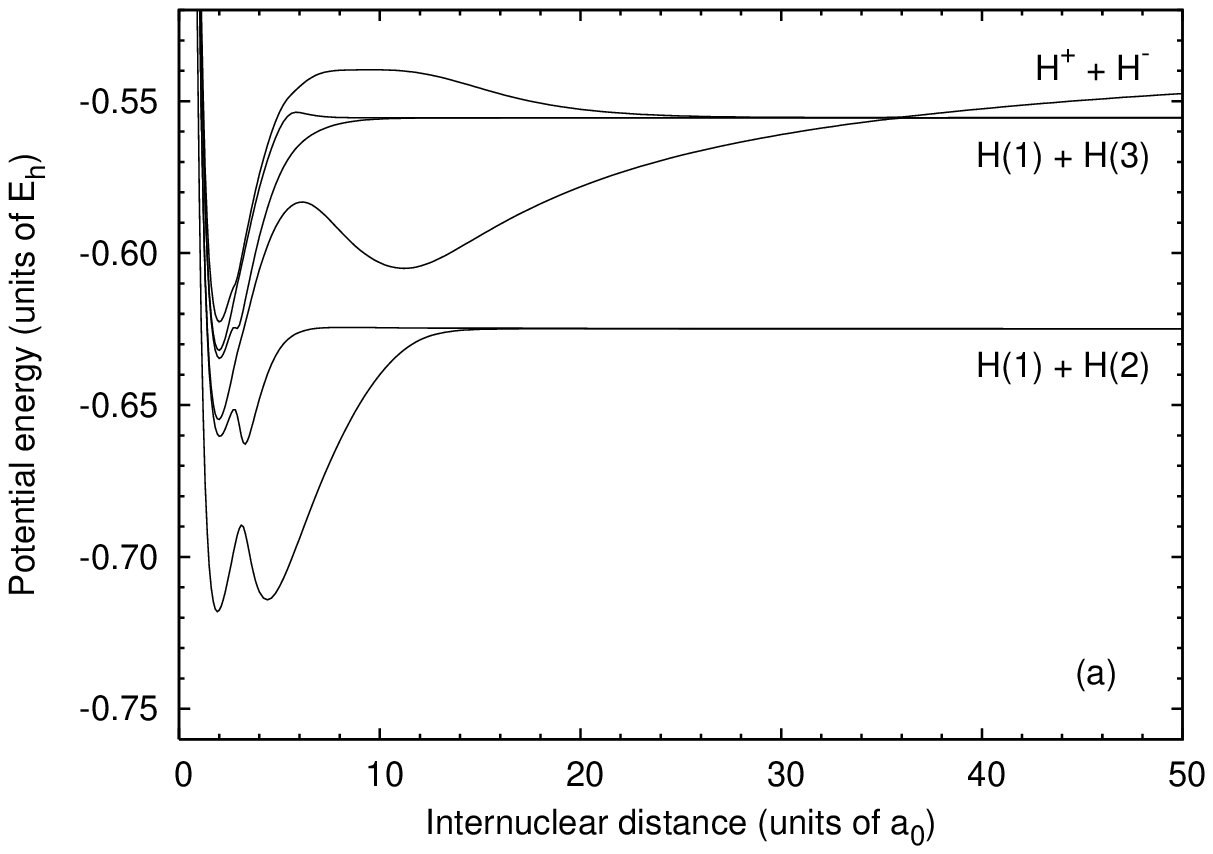}}
\hfill
\scalebox{0.68}{\includegraphics{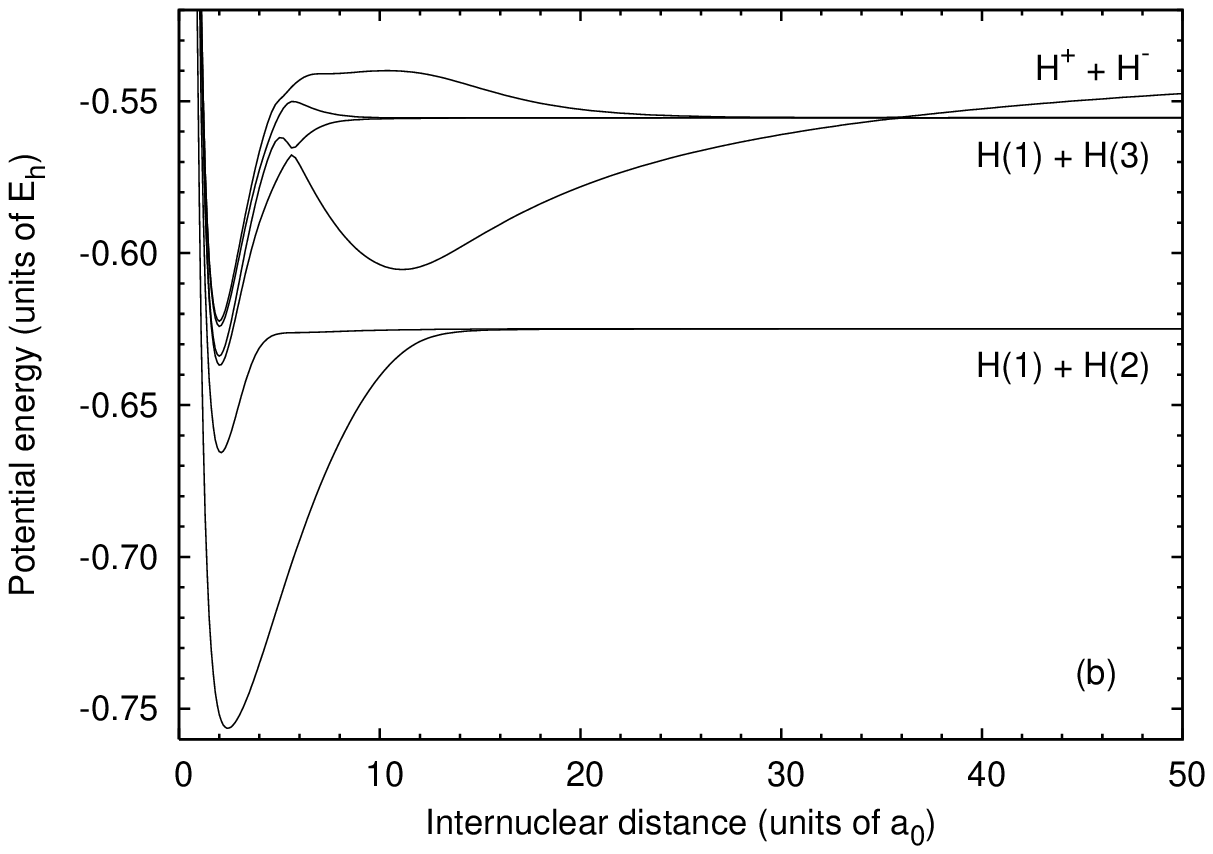}}
\caption{\label{figure:pecs}
Adiabatic potential energy curves for (a) the (2-7)$^1\Sigma_g^+$ states
and (b) the (1-6)$^1\Sigma_u^+$ states of $\text{H}_2$. The separated atomic limits
are indicated at the far right of each subfigure. For a discussion of the particular
ion-pair limit, see the main text.}
\end{figure}
    
Figure~\ref{figure:pecs} shows the (2-7)$^1\Sigma_g^+$ and (1-6)$^1\Sigma_u^+$
potential energy curves in the range 0.5 to 50~$\text{a}_0$. As is well known, only one
bound state of the negative hydrogen ion exists; the $^1\text{S}^e$ ground
state~\cite{rau96}. Accordingly, there is one gerade and one ungerade electronic state
which asymptotically correlate with the $\text{H}^++\text{H}^-$ ion-pair limit. As the
internuclear distance is decreased from infinity, the ion-pair configuration is successively
carried on through a series of avoided crossings involving states correlated with the
$\text{H}(1)+\text{H}(n)$ covalent limits. In the neighborhood of these crossings, the
first derivative radial couplings~(\ref{eq:radnac}) are significant and are able to cause
transitions between different adiabatic states.
      
Beginning with the gerade symmetry, the (1-6)$^1\Sigma_g^+$ manifold consists of
states correlated with the $n=1,2,3$ covalent limits. Corresponding to each limit
$\text{H}(1)+\text{H}(n)$ there are $n$ of these states in total. The 7$^1\Sigma_g^+$
state has the ion-pair configuration from $36~\text{a}_0$ up to approximately
$280~\text{a}_0$. Here, the adiabatic wave function changes character into the $n=4$
covalent configuration while the ion-pair configuration is carried on to another higher
excited state. At such large distances, however, the motion of the nuclei is highly diabatic
and the change of character of the adiabatic wave functions will not induce any significant
ionic-covalent transitions~\cite{bat55,sid83,bor83}. With these considerations in mind,
we will refer to the 7$^1\Sigma_g^+$ state as that which is correlated with the ion-pair
limit.

In the ungerade symmetry, the (1-5)$^1\Sigma_u^+$ manifold consists of states
correlated with the $n=2,3$ covalent limits while the 6$^1\Sigma_u^+$ state, with the
same considerations as above, can be said to correlate with the ion-pair limit. Because of
symmetry restrictions, there can be no singlet ungerade state dissociating into the $n=1$
covalent limit. Note that while the gerade and ungerade states behave very
differently at small internuclear distances, these differences vanish quickly as the
separation of the nuclei is increased. In fact, at $10~\text{a}_0$ the energy difference
for all excited states is less than $9\times10^{-4}~E_\text{h}$ (Hartree) and at
$30~\text{a}_0$ it is less than $5\times10^{-5}~E_\text{h}$. 
      
Figure \ref{figure:pecs} illustrates the two avoided crossings near 12 and $36~\text{a}_0$,
which are related to a change from ionic to $n=2$ and 3 covalent character (or vice versa)
of the adiabatic wave functions, respectively. In what follows we will refer to them simply
as the $n=2$ and 3 curve crossings. Figure~\ref{figure:pecs} also shows a rich pattern of
avoided crossings taking place at smaller internuclear distances. This region, however, is
to a large extent shielded by the centrifugal term in Eq.~(\ref{eq:rse1}) and the influence of the
resulting radial couplings upon the total neutralization cross section is rather small (see
Sec.~\ref{level2:tcs}). The higher excited $^1\Sigma_{g,u}^+$ states that are not considered in the
present study are only accessible to the system through avoided crossings at small internuclear
distances~\cite{det98}. The error introduced by excluding these states is therefore expected to be of
less significance. In principle, also rotational couplings to states of $^1\Pi$ symmetry should be
considered. However, due to the presence of strong radial couplings in the vicinity of the $n=2$ and 3
curve crossings, and due to the low collision energies considered in the present study, these are
assumed to be negligible~\cite{fus86,dic99}.
      
To evaluate the quality of our calculated potential energy curves, we have compared them
with those reported by Wolniewicz and co-workers. Their calculations cover the
1$^1\Sigma_g^+$ state out to $12~\text{a}_0$~\cite{wol93,wol95}, the
(2-6)$^1\Sigma_g^+$ states out to $20~\text{a}_0$~\cite{wol94} and the
(1-6)$^1\Sigma_u^+$ states out to $150~\text{a}_0$~\cite{wol02}. The potential
energy curve for the 4$^1\Sigma_g^+$ state has been recalculated and extended out to
$80~\text{a}_0$~\cite{wol98}. With the energy scale used in Fig.~\ref{figure:pecs}, these
potentials would entirely overlap with ours. A more detailed comparison shows that the
largest energy difference is obtained for the inner part ($R\le1.5~\text{a}_0$) of the
1$^1\Sigma_g^+$ ground state potential, which is of little importance to the present
study anyway. Here, our computed values are approximately $5.7\times10^{-4}$ to
$1.5\times10^{-3}~E_\text{h}$ higher in energy compared to those in Ref.~\cite{wol95}. For
all of the excited states, our calculated energies are approximately $1.0\times10^{-4}$ to
$5.6\times10^{-4}~E_\text{h}$ higher than in Refs.~\cite{wol94,wol98,wol02}. It is of
particular interest to estimate the quality of the (4-7)$^1\Sigma_g^+$ and
(3-6)$^1\Sigma_u^+$ states near the critical $n=3$ curve crossing. Concerning the
ungerade states, the covalent parts of our potentials are within
$1.6\times10^{-4}~E_\text{h}$ and the ionic parts within $3.0\times10^{-4}~E_\text{h}$ to
those reported in Ref.~\cite{wol02}. Due to the near degeneracy of the gerade and ungerade states,
we expect this accuracy to be common to both inversion symmetries. The quality of the
adiabatic states and in particular their energy splitting at the $n=3$ curve crossing will
be discussed further in the next section.

\subsection{\label{level2:rcoups}Radial couplings}
We have calculated the first derivative radial couplings~(\ref{eq:radnac}) between all of
the adiabatic states considered in the previous section. The accurate evaluation of these
quantities is not trivial. The magnitude and shape of the radial couplings are usually very
sensitive to the quality of the electronic wave functions and it is essential that not
only the individual states are good in a variational sense, but also that the relative
energies of these states are well represented. In regions where two or more adiabatic
states become nearly degenerate, further complexity is added by the fact that a minor
change in the atomic basis can cause the potential energy curves to artificially pseudo
cross, giving rise to dramatic effects in the radial couplings. These effects do not
change the outcome of the dynamics but can obscure both interpretation and comparison with
other results.

To evaluate the radial couplings we have implemented a three point version of the finite
difference method described in Ref.~\cite{gal77}. This method provides a systematic way to
converge the calculations toward the exact result in the particular type of basis and wave
function being considered. We have examined carefully how the calculated radial couplings
depend on the derivative step length $\Delta{R}$, as well as the numerical accuracies,
$\delta_\text{MO}$ and $\delta_\text{CI}$, with which the molecular orbitals and the CI
wave functions are obtained. Stable and converged results were observed when
$\Delta{R}=5\times10^{-5}~a_0$, $\delta_\text{MO}=10^{-14}~E_\text{h}$ and
$\delta_\text{CI}=10^{-13}~E_\text{h}$. The antisymmetry condition~(\ref{eq:asc}) can be
used as a simple consistency check of the results. In the present case, most of the calculated
coupling elements satisfied this condition to at least four significant digits. For practical
reasons, it is still desirable to have a coupling matrix that is fully antisymmetric.
This has been accomplished by taking each element to be the geometrical mean of
$\tau_{ij}$ and $-\tau_{ji}$.
      
The radial couplings are too many to discuss in full detail and so we will focus
on those relevant for the important $n=2$ and 3 curve crossings.
The notation $(i,j)_g$ is used to label the radial coupling between the $i$-th
and $j$-th $^1\Sigma_g^+$ state (and similarly for the ungerade states). Due to
symmetry, gerade and ungerade states are not coupled to each other.  
\begin{figure}[htp]
\scalebox{0.68}{\includegraphics{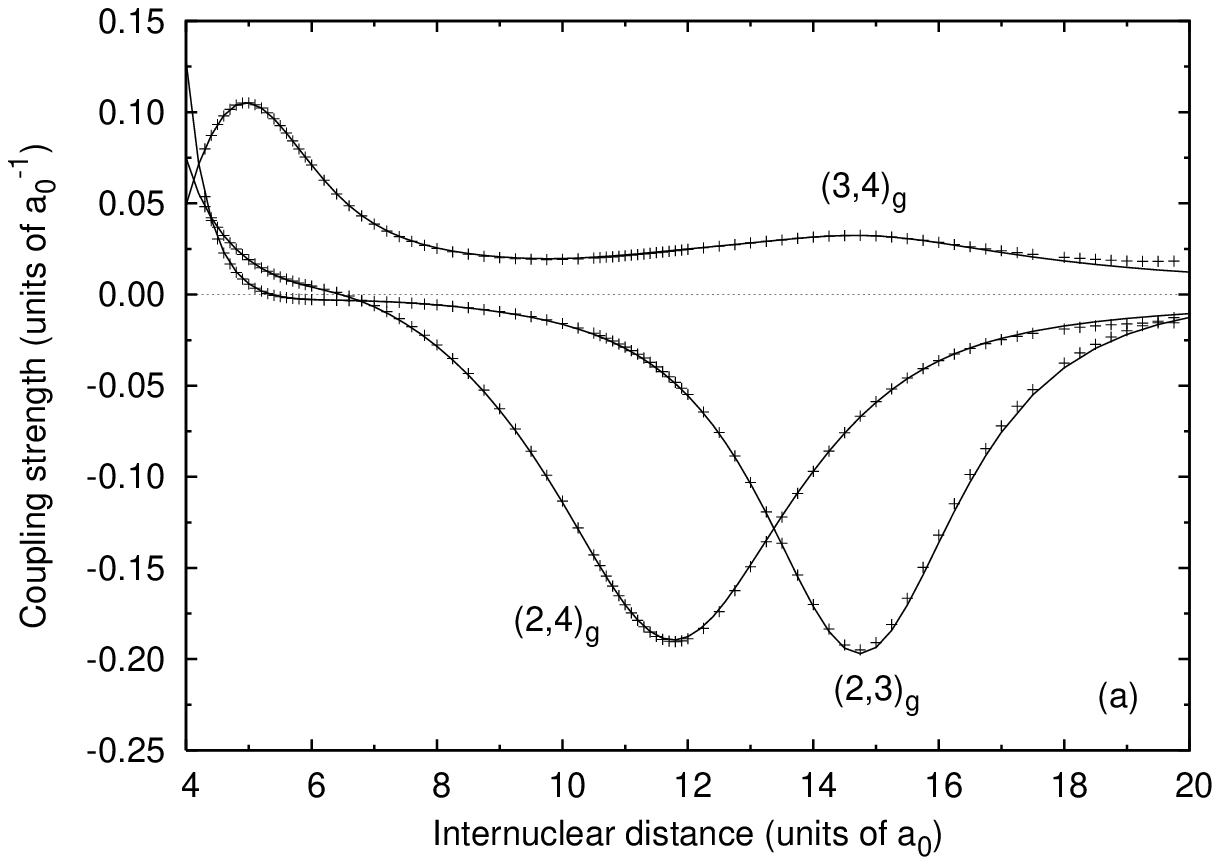}}
\hfill
\scalebox{0.68}{\includegraphics{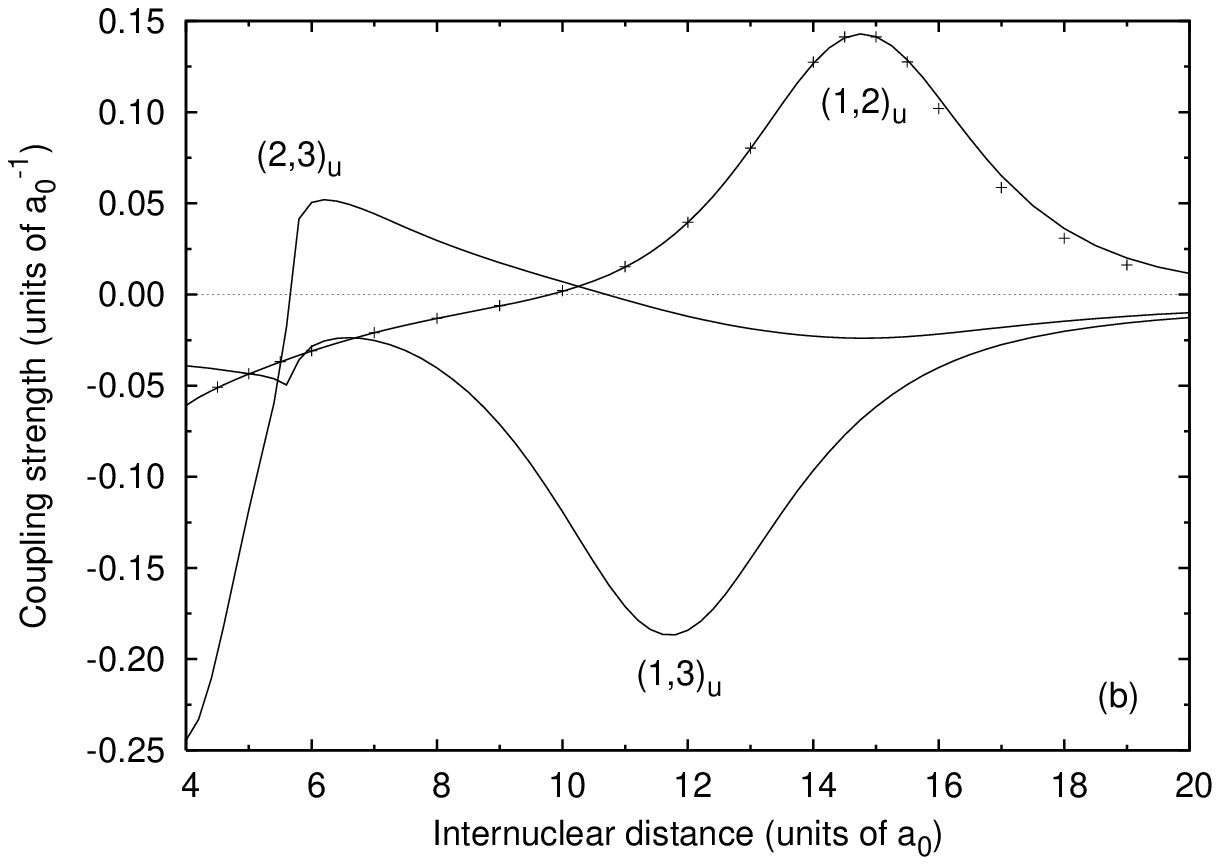}}
\caption{
\label{figure:coups2}
Radial couplings among (a) the (2-4)$^1\Sigma_g^+$ states and (b) the
(1-3)$^1\Sigma_u^+$ states in the neighborhood of the $n=2$ curve crossing. The
notation $(i,j)_g$ label the radial coupling between the $i$-th and $j$-th
$^1\Sigma_g^+$ state (and similarly for the ungerade states). The present
results (solid lines) are compared with the results obtained by Wolniewicz and
Dressler (crosses)~\cite{wol88,wol94}.}
\end{figure}
      
In Fig.~\ref{figure:coups2} we show the radial couplings among the
(2-4)$^1\Sigma_g^+$ and (1-3)$^1\Sigma_u^+$ states in the neighborhood of
the $n=2$ curve crossing. As can be expected from the appearance of the potential energy
curves, each of the $(2,4)_g$ and $(1,3)_u$ couplings show a rather broad peak
near the crossing point at $12~a_0$ resulting from the exchange of ionic and $n=2$
covalent character between the adiabatic wave functions. Similar shapes and magnitudes,
although displaced to larger internuclear distances, are also exhibited by the
$(2,3)_g$ and $(1,2)_u$ couplings, which connect the adiabatic states tending
to the $n=2$ covalent limit. The $(3,4)_g$ and $(2,3)_u$ couplings are, on the
other hand, much less pronounced in the crossing region. For comparison, also plotted
in Fig.~\ref{figure:coups2} are the results obtained
by Wolniewicz and Dressler~\cite{wol88,wol94}, which include the radial
couplings among all of the (2-4)$^1\Sigma_g^+$ states and between the 1 and
2$^1\Sigma_u^+$ states. The overall agreement with the present results is very good, in
particular in the region up to approximately $16~a_0$. For larger internuclear distances
some deviations can be observed. These are most likely due to a slight unbalance in our
description of the nearly degenerate $n=2$ adiabatic states. However, as mentioned at the
very beginning of this section, the physical relevance of these types of effects is
expected to be small.
\begin{figure}[htp]
\scalebox{0.68}{\includegraphics{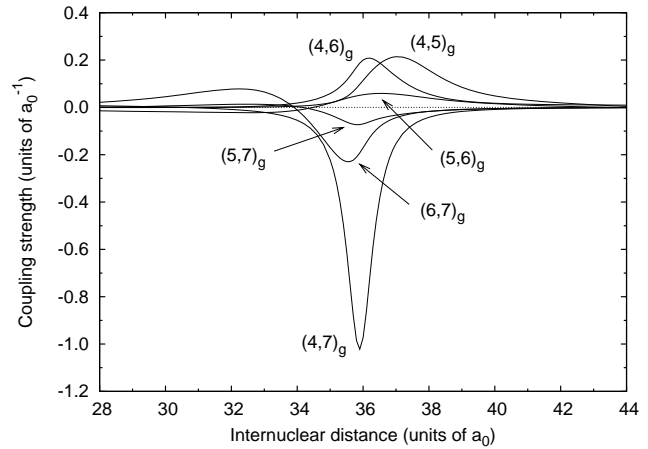}}
\caption{
\label{fig:coups3}
Radial couplings among the (4-7)$^1\Sigma_g^+$~states in the neighborhood of the $n=3$
curve crossing. The notation $(i,j)_g$ label the radial coupling between the $i$-th
and $j$-th $^1\Sigma_g^+$ state.}
\end{figure}

Figure~\ref{fig:coups3} illustrates the radial couplings among the adiabatic states associated
with the $n=3$ curve crossing. Here, the gerade and ungerade states are almost completely
degenerate and so only the results for the former of these, i.e., the (4-7)$^1\Sigma_g^+$
manifold of states, are shown. All the couplings
appear as peak like structures with their maxima located in the region 35.5 to $37.0~a_0$.
The most prominent of these is the $(4,7)_g$ coupling which reaches its maximum value
at $35.9~a_0$. This coupling arises from the exchange of ionic and $n=3$
covalent character between the 4 and 7$^1\Sigma_g^+$ adiabatic states. We note that
the location of its maximum agrees well with the crossing point of the pure ionic and
covalent energies at $36.0~a_0$ obtained from the ground state energy~\cite{tha94} and the
polarizability~\cite{hol05} of the $\text{H}^-$ ion. The calculations of Wolniewicz and
Dressler do not extend out to the $n=3$ curve crossing and thus no direct comparison of
the radial couplings can be made. However, in the ungerade symmetry we can compare the
energy splitting of the 3 and 6$^1\Sigma_u^+$ states at $35.9~a_0$. Here, we obtain
the value $3.8\times10^{-4}~E_\text{h}$ which is in very good agreement with the value
$3.9\times10^{-4}~E_\text{h}$ calculated from the results of Ref.~\cite{wol02}.

\subsection{\label{level2:atdt}Adiabatic to diabatic transformation}
\begin{figure}[htp]
\scalebox{0.68}{\includegraphics{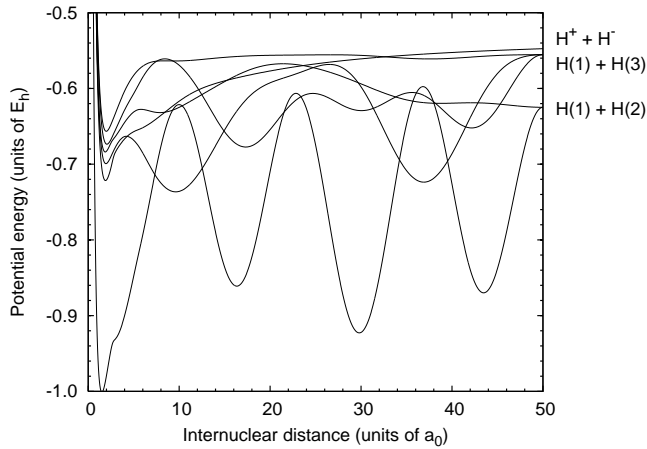}}
\caption{
\label{figure:potsdg}
Strictly diabatic potential energy curves obtained from a transformation of the
(1-7)$^1\Sigma_g^+$ adiabatic states. The diabatic potential energy curve
associated with the $\text{H}(1)+\text{H}(1)$ limit is not shown in the figure.}
\end{figure}

In order to obtain the adiabatic to diabatic transformation (ATDT) matrix, we have
numerically solved Eq.~(\ref{eq:atdte2}) using the radial couplings described in the preceding
section as input data. The solution has been obtained with a matrix version of the
Runge-Kutta Fehlberg method~\cite{che85}. The boundary condition $\mathbf{T}=\mathbf{1}$
has been imposed at $R=50~a_0$ and the integration has been performed inwards. As hinted
above, several of the non-adiabatic couplings are non-vanishing as $R$ tends to infinity. In
practice, our choice of boundary condition corresponds to setting these couplings to zero
beyond $50~a_0$. Since this kind of assumption has to be made in the calculation of the
scattering matrix anyway, no additional error is introduced due to the diabatization
procedure itself. In the present case, the transformation to a strictly
diabatic basis is merely for computational reasons and is not discussed further. It is worth noting,
however, that the diabatic potential energy curves we obtain here are far
from the intuitive ones conventionally used in connection with the $\text{H}_2$ system, see
Figure~\ref{figure:potsdg}.

\section{\label{level1:numpro}Numerical procedures}      
The logarithmic derivative~(\ref{eq:ydef}) has been calculated by numerically
solving the matrix Riccati equation~(\ref{eq:riceq}) from $0.54~\text{a}_0$
out to $50~\text{a}_0$. For this purpose we have used an algorithm developed by
Johnson~\cite{joh73,man93} with the grid size set to $5\times10^{-3}~\text{a}_0$. Knowledge
of the logarithmic derivative in the final integration point has allowed us to calculate the
partial cross sections~(\ref{eq:partcs}) for scattering within the gerade and ungerade
inversion symmetries. The fully symmetrized integral cross sections have been obtained
by combining the partial cross sections according to Eq.~(\ref{eq:h2cs}). To determine at
which point the series in Eq.~(\ref{eq:h2cs}) could be truncated, a simple convergence
criteria was introduced. This was set to terminate the summation if the ratios of the
partial cross sections and the accumulated integral cross sections remained less than
$10^{-4}$ for 25 terms in succession. Over the energy range considered here, this led to the
inclusion of approximately 250 to 3500 partial waves.   

\section{\label{level1:res}Results and discussion}

\subsection{\label{level2:stscs}State dependent cross sections}     
\begin{figure}[htp]
\scalebox{0.68}{\includegraphics{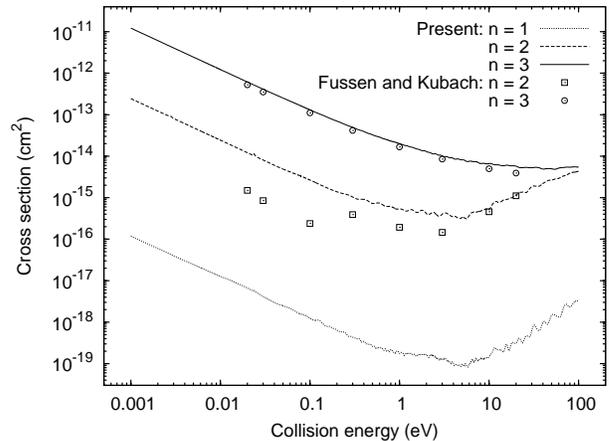}}
\caption{
\label{figure:crossstate}
Calculated cross sections for scattering into the $\text{H}(1)+\text{H}(n)$, $n=1,2,3$
final states. The present results are compared with the close-coupling results of Fussen
and Kubach~\cite{fus86}.}
\end{figure}
 
The calculated cross sections for scattering into the $\text{H}(1)+\text{H}(n)$,
$n=1,2,3$ final states are shown in Fig.~\ref{figure:crossstate}. It is clear that the $n=3$
process dominates the others over almost the entire energy range, which is reasonable
considering the favorable location of the $n=3$ curve crossing at around $36~a_0$. In
contrast, the $n=2$ transitions occur much further in and are more easily suppressed by the
centrifugal barrier. Only for collision energies above $5$~eV, where the $n=2$ cross section
exhibits a minimum, is this process able to compete with neutralization into the $n=3$
final states. This phenomena was first noted by Bates and Lewis~\cite{bat55} and has been
confirmed, at least on a qualitative level, by almost every study since
then~\cite{ols70,eer95,fus86}. Neutralization into the $n=1$ final state is insignificant at
all energies and is included only for completeness. As the collision energy approaches zero,
all cross sections gain the characteristic $E^{-1}$ dependence that can be expected for
reactions which are governed by the Coulomb interaction. Here, the $n=3$ cross section is
approximately a factor of~50 larger than the $n=2$ cross section.

Some weaker oscillations can be observed over a large part of the energy range. The reason for this
structure is not completely clear, but the most likely explanation is in terms of
St{\"u}ckelberg oscillations, i.e., in terms of quantum interference arising because there
are several competing routes through the potential landscape leading to the same asymptotic limit.
      
The close-coupling results of Fussen and Kubach~\cite{fus86} have also been plotted in
Fig.~\ref{figure:crossstate} for comparison.
The agreement between their $n=3$ cross section
and the present one is good. The difference is about 20~percent at 1~eV and
increases slightly with the energy. Comparing the $n=2$ cross sections, the two calculations
are seen to agree well above the minimum at 5~eV, whereas at lower energies the differences
are more pronounced. A rationale for this discrepancy may be found by considering how the
inversion symmetry of the electronic states is approached.
\begin{figure}[htp]
\scalebox{0.68}{\includegraphics{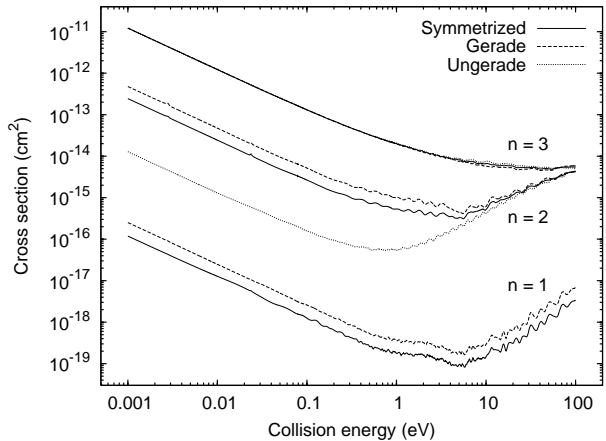}}
\caption{
\label{figure:crossgu}
Fully symmetrized vs. pure gerade and ungerade cross sections for
scattering into the $\text{H}(1)+\text{H}(n)$, $n=1,2,3$ final states.}
\end{figure}
For this purpose we show in Fig.~\ref{figure:crossgu} a comparison of our fully symmetrized
cross sections (same as in Fig.~\ref{figure:crossstate}) together with the pure gerade and
ungerade cross sections calculated in the conventional way. We first note that the $n=3$
cross section is fairly insensitive to the choice of inversion symmetry over the whole
energy range. Consequently, the effect of averaging over these symmetries is expected to be small.
A similar observation can be made regarding the $n=2$ cross section above 5~eV. However, at lower
energies the gerade and ungerade results differ strongly. This indicates not only that
our way of combining the gerade and ungerade cross sections will affect the final result, but also
that any two approaches that treat the inversion symmetry differently may lead to different outcomes.
In the one electron model of Fussen and Kubach, this inversion symmetry is actually broken and
as a consequence their results are likely to differ from ours. It should be emphasized
though, that at the energies where this happens the contribution from the $n=2$ channels
to the overall reaction is small. It should also be noted that the effects of treating the
nuclei as indistinguishable are very small, and it is equally good to use the simpler
formula~(\ref{eq:h2cs2}) in favor of (\ref{eq:h2cs}) to weight the gerade and ungerade partial
cross sections.

\subsection{\label{level2:tcs}Total neutralization cross section}
\begin{figure*}[htp]
\scalebox{1.4}{\includegraphics[clip=true, viewport=0.1cm 3.6cm 12.5cm 8.65cm]{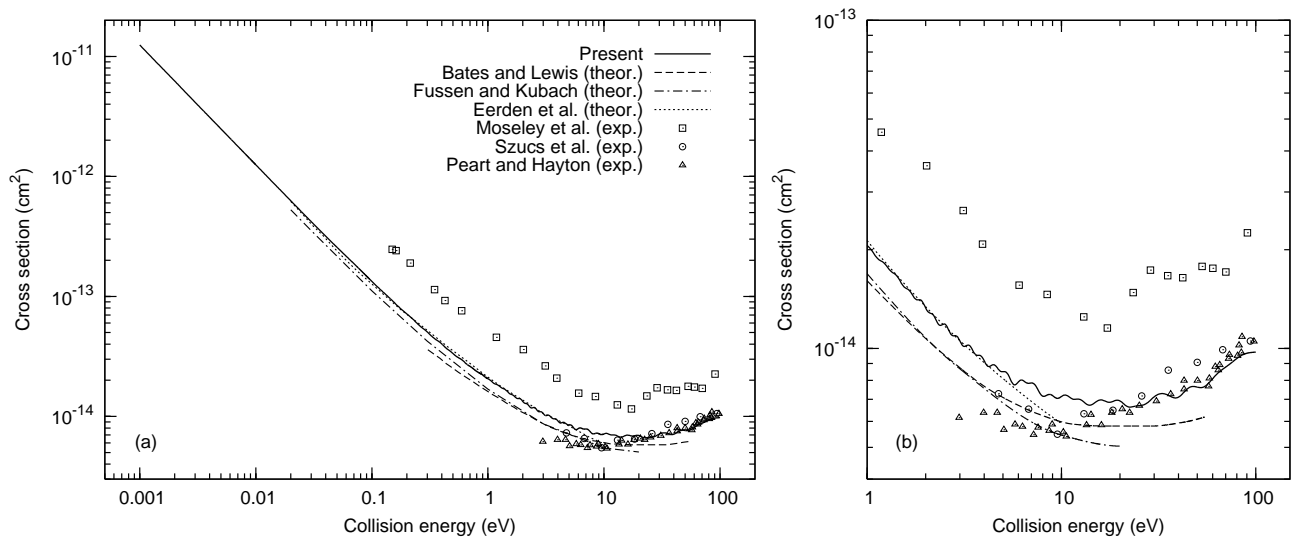}}
\caption{
\label{figure:crosstot}
Total cross section for the mutual neutralization of $\text{H}^+$ and $\text{H}^-$. The
present results are compared with the measurements by Moseley \textit{et al.}~\cite{mos70},
Szucs \textit{et al.}~\cite{szu84}, and Peart and Hayton~\cite{pea92}, as well as the
calculations by Bates and Lewis~\cite{bat55}, Fussen and Kubach~\cite{fus86}, and Eerden
\textit{et al.}~\cite{eer95}. (a) Cross sections plotted over the full energy range 0.001 to 100 eV.
(b) Enlargement of the energy range 1 to 100 eV.}
\end{figure*}

In Fig.~\ref{figure:crosstot} we show the calculated total neutralization cross section,
i.e., the sum of the $n=1,2,3$ cross sections. As could be expected from the previous subsection,
this cross section shows a rather smooth behavior over the entire energy range. At low
collision energies, the curve approximately resembles the $E^{-1}$ behavior discussed
above. The cross section then continues to fall off until it reaches a broad minimum at around
20 eV. This minimum obviously results from the nearly constant $n=3$ cross section in combination
with the rapid increase in the formation of $n=2$ final products.

To examine the influence of the non-vanishing asymptotic couplings upon the cross section,
the point at which the logarithmic derivative is evaluated has been varied over the range 50
to $80~\text{a}_0$. The boundary condition on the ATDT matrix has been varied accordingly.
At the collision energies 0.001, 1 and 100 eV, the relative variation in the cross
section was seen to be $4\times10^{-4}$, $7\times10^{-4}$ and $1\times10^{-3}$,
respectively. We conclude that the effects of the non-vanishing asymptotic couplings over
the energy range considered here are negligible. This is consistent with the conclusions drawn by
Borondo \textit{et al.}~\cite{bor83}.

Additional tests have been performed in order to estimate the relevance of the radial couplings at
small internuclear distances (as discussed in Sec.~\ref{level2:pes}). This has been done by
setting all radial couplings up to a certain distance~$R_0$ to zero before the diabatization
procedure is carried out. Choosing $R_0=6~\text{a}_0$ reduces the total
neutralization cross section at 0.001, 1 and 100 eV collision energy by 8.6, 9.9 and 5.9 percent,
respectively. These variations in the cross section can be considered small (although not
insignificant) and may be taken as an upper limit for the influence of the higher lying $^1\Sigma_{g,u}^+$
adiabatic states not included in the present study, which are coupled to the lower lying states
mainly at small internuclear distances. A more detailed study of these effects is currently
being pursued.

In Fig.~\ref{figure:crosstot} we also show the results of earlier studies relevant for the
mutual neutralization of $\text{H}^+$ and $\text{H}^-$ at low collision energies. Up to the
present date, no measurements of this cross section have been reported for collision
energies below 0.15~eV. In the region 0.15 to 3~eV, the only experimental results that are
available are those of Moseley \textit{et al.}~\cite{mos70}. A comparison shows that these
are a factor of two to three larger than the present data. At energies above a few~eV,
experiments have been conducted by Szucs~\textit{et al.}~\cite{szu84} and Peart and
Hayton~\cite{pea92}, among others. In the region 3 to 10~eV, their results point towards a
cross section that is slightly lower than ours, which is most evident for the first few data
points of Peart and Hayton, whereas for collision energies above 10~eV all three studies agree
favorably. Comparing with earlier theoretical studies, our results match within 30 percent the
close-coupling calculations of Fussen and Kubach~\cite{fus86} and within 40 percent the
Landau-Zener calculations of Bates and Lewis~\cite{bat55}. Our results also more or less
overlap with the Landau-Zener results obtained by Eerden \textit{et al.}~\cite{eer95}.
      
It may be worthwhile to review the accumulated data on the mutual neutralization of
$\text{H}^+$ and $\text{H}^-$ at low collision energies (below a few eV). There are now
several calculations of this cross section that are mutually consistent
within a few tens of percent. These employ such diverse methods as semi-classical
Landau-Zener theory as well as model Hamiltonian and \textit{ab initio} close-coupling
schemes. In the upper part of the energy region, the theoretical results are in more
or less agreement with several of the available experiments. One of these might even suggest
a cross section that is slightly lower than predicted by theory. It is a fact, however, that
the only measurement that has been published for collision energies below 3~eV shows a
cross section that is considerably higher than others reported. To clarify this situation,
new experimental efforts are obviously required. Recently such measurements have been
undertaken by the group of Urbain~\cite{urb08} at the Universit{\'e} Catholique de Louvain,
Belgium, using a merged beam apparatus in the energy range 0.001 to 0.2~eV. Although
uncertainties of the order of 50 percent still exist, preliminary results from this
experiment seem to support the neutralization cross section obtained from the various
theoretical treatments.

\subsection{\label{rcf:level2}Rate coefficient}     
At low collision energies, the total neutralization cross section may be parametrized in
terms of the relative velocity~$v$ as~\cite{mos70}
\begin{equation}
\label{eq:rc}
\sigma(v)=Av^{-2}+Bv^{-1}+C+Dv~.
\end{equation}
To determine the unknown coefficients of this expression, we have performed a least square
fit to our calculated cross section in the region 0.001 to 4~eV. The result is
\begin{eqnarray}
A&=&4.77\times10^{-2}~\text{cm}^{4}\text{s}^{-2}~,\nonumber\\
B&=&-1.73\times10^{-9}~\text{cm}^{3}\text{s}^{-1}~,\nonumber\\
C&=&1.22\times10^{-14}~\text{cm}^{2}~,\nonumber\\
D&=&-1.57\times10^{-21}~\text{cm}~\text{s}~,\nonumber
\end{eqnarray}
corresponding to a maximum error of six percent in the parametrized cross section. By integrating
Eq.~(\ref{eq:rc}) over a Maxwellian velocity distribution, the associated rate 
coefficient~$\alpha(T)$ is obtained:
\begin{eqnarray}
\alpha(T)&=&A\left(\frac{2\mu}{\pi{k}T}\right)^{1/2}+B\nonumber\\
&&+2C\left(\frac{2kT}{\pi\mu}\right)^{1/2}+\frac{3DkT}{\mu}~,
\end{eqnarray}
where $T$ is the ion temperature and $k$ the Boltzmann constant. We estimate the above
results to be valid over the temperature range 10 to 10,000~K. For comparison, the rate
coefficient we obtain at 300 K is $1.73\times10^{-7}~\text{cm}^{3}\text{s}^{-1}$.

\section{\label{level1:sum}Summary}
In this paper we have considered the mutual neutralization of $\text{H}^++\text{H}^-$ into
$\text{H}(1)+\text{H}(n)$, $n=1,2,3$ at low collision energies. To arrive at the present
results, we have used a molecular close-coupling approach with all degrees of freedom
treated quantum mechanically. Adiabatic potential energy curves and non-adiabatic radial
couplings have been calculated at the FCI level of theory employing a large Gaussian basis
set. These quantities are in good agreement with those obtained from more sophisticated
electronic structure methods~\cite{wol88,wol90,wol93,wol94,wol95,wol98,wol02}. Using a
strictly diabatic representation of the potential energy curves and coupling matrix elements,
we have calculated state dependent and total neutralization cross sections taking
into account the identity of the nuclei.

The present results conform to the conventional view that for collision energies up to a
few eV, almost all neutral products go into the $n=3$ final states. Furthermore, 
a proper weighting of the gerade and ungerade cross sections turns out to be important only
for the $n=2$ formation at low collision energies, where these cross sections anyway are small.
In the collision energy region below a few eV, our total neutralization cross section is in good
agreement with previous theoretical studies~\cite{bat55,fus86,eer95}, but is a factor of
two to three lower than that measured by Moseley \textit{et al.}~\cite{mos70}.

\begin{acknowledgments}
This work was supported by the Swedish Research Council (VR). We thank X. Urbain and A.~E.
Orel for useful discussions, and R.~D. Thomas for reading and commenting on the manuscript.
\end{acknowledgments}

\appendix*
\section{}
Here, we provide analytical expressions for the solution matrices $\mathbf{a}_l$,
$\mathbf{b}_l$, $\bm{\alpha}_l$ and $\bm{\beta}_l$ which are used in the definitions of the
reactance and scattering matrices of Sec.~\ref{level2:scafor}. Open and closed channels
are distinguished depending on whether the total energy is greater or less than the
corresponding threshold energy. The channels are labeled short-range (covalent) if
\begin{equation}
\label{eq:cov}
V_{ii}^\text{d}(R)\underset{R\ge\rho}{=}E_i^\text{th}
\end{equation}
and Coulombic if
\begin{equation}
\label{eq:coul}
V_{ii}^\text{d}(R)\underset{R\ge\rho}{=}E_i^\text{th}+\frac{Q_1Q_2}{R}~,
\end{equation}
where $Q_1$ and $Q_2$ are the net charges situated on each of the nuclei. The asymptotic
wave number $k_i$ for open channels has been defined in Eq.~(\ref{eq:k1}). For closed
channels,
\begin{equation}
\label{eq:k2}
k_i=\left[2\mu\left(E_i^\text{th}-E\right)\right]^{1/2}~.
\end{equation}
It is further convenient to introduce the scaled coordinate $\xi_i=k_iR$ and the Coulomb
parameter
\begin{equation}
\label{eq:coupar}
\eta_i= \frac{\mu Q_1Q_2}{k_i}.
\end{equation}

We look for pairs of linearly independent solutions to the radial Schr{\"o}dinger
equation (\ref{eq:rse2}) in the region $R\ge\rho$. These are chosen such that
$a_{ii,l}(R)$ and $b_{ii,l}(R)$ behave as regular and irregular solutions, and
$\alpha_{ii,l}(R)$ and $\beta_{ii,l}(R)$ as incoming and outgoing wave solutions,
respectively. For open short-range channels, two linearly independent solutions are the
Riccati-Bessel functions of the first and second kind, $\widetilde{S}_l(\xi_i)$ and
$\widetilde{C}_l(\xi_i)$, respectively~\cite{abr72}. We adopt the definitions~\cite{joh85}
\begin{equation}
a_{ij,l}=\delta_{ij}k_i^{-1/2}\widetilde{S}_l~,
\end{equation}
\begin{equation}
b_{ij,l}=\delta_{ij}k_i^{-1/2}\widetilde{C}_l~,
\end{equation}
\begin{equation}
\alpha_{ij,l}=\delta_{ij}k_i^{-1/2}\left(-i\widetilde{S}_l+\widetilde{C}_l\right)~,
\end{equation}
\begin{equation}
\beta_{ij,l}=\delta_{ij}k_i^{-1/2}\left(i\widetilde{S}_l+\widetilde{C}_l\right)~.
\end{equation}
For open Coulomb channels, the radial Schr\"odinger equation is solved by the regular and
irregular Coulomb functions, $\widetilde{F}_l(\eta_i,\xi_i)$ and
$\widetilde{G}_l(\eta_i,\xi_i)$, respectively~\cite{abr72}. Accordingly,
\begin{equation}
a_{ij,l}=\delta_{ij}k_i^{-1/2}\widetilde{F}_l~,
\end{equation}
\begin{equation}
b_{ij,l}=\delta_{ij}k_i^{-1/2}\widetilde{G}_l~,
\end{equation}
\begin{equation}
\alpha_{ij,l}=\delta_{ij}k_i^{-1/2}\left(-i\widetilde{F}_l+\widetilde{G}_l\right)~,
\end{equation}
\begin{equation}
\beta_{ij,l}=\delta_{ij}k_i^{-1/2}\left(i\widetilde{F}_l+\widetilde{G}_l\right)~.
\end{equation}
Finally, in the case of closed short-range channels, possible solutions are
$\xi_i\widetilde{i}_l(\xi_i)$ and $\xi_i\widetilde{k}_l(\xi_i)$, where
$\widetilde{i}_l(\xi_i)$ and $\widetilde{k}_l(\xi_i)$ are the modified spherical Bessel
functions of the first and second kind, respectively~\cite{abr72} (first and third kind
in the cited reference). Appropriate definitions are~\cite{joh85}
\begin{equation}
a_{ij,l}=\delta_{ij}2^{1/2}\pi^{-1/2}\xi_i\widetilde{i}_l~,
\end{equation}
\begin{equation}
b_{ij,l}=\delta_{ij}2^{1/2}\pi^{-1/2}\xi_i\widetilde{k}_l~,
\end{equation}
\begin{equation}
\alpha_{ij,l}=\delta_{ij}(2\pi)^{1/2}\xi_ik_i^{-1}
\left[\widetilde{i}_l+(-1)^l\pi^{-1}\widetilde{k}_l\right]~,
\end{equation}
\begin{equation}
\beta_{ij,l}=\delta_{ij}2^{1/2}\pi^{-1/2}\xi_ik_i^{-1}\widetilde{k}_l~.
\end{equation}
Closed Coulomb channels play no role in the present study and are therefore not considered.

\end{document}